\documentclass[prb,amsmath,amssymb,floatfix,twocolumn,citeautoscript]{revtex4}
\usepackage{bm}
\usepackage{epsfig}
\usepackage{subfigure}
\usepackage[usenames]{color}

\begin{document}

\title{Anomalous Scaling of the Penetration Depth in Nodal Superconductors}

\author{Jian-Huang She$^1$, Michael J. Lawler$^{1, 2}$, and Eun-Ah Kim$^1$}

\affiliation{$^1$Department of Physics, Cornell University, Ithaca, New York 14853, USA \\
$^2$Department of physics, Binghamton University, Vestal NY 13850}

\begin{abstract}

Recent findings of anomalous super-linear scaling of low temperature ($T$) penetration depth (PD) in several nodal superconductors near putative quantum critical points 
suggest that the low temperature PD can be a useful probe of quantum critical fluctuations in a superconductor. On the other hand, cuprates which are poster child nodal superconductors 
have not shown any such anomalous scaling of PD, despite growing evidence of quantum critical points. 
Then it is natural to ask when and how can quantum critical fluctuations cause anomalous scaling of PD?  Carrying out the renormalization group calculation for the problem of two dimensional superconductors with point nodes, we show that quantum critical fluctuations associated with point group symmetry reduction result in non-universal logarithmic corrections to the $T$-dependence of the PD.  The resulting  apparent power law depends on the bare velocity anisotropy ratio. We then compare our results to data sets from two distinct nodal superconductors: YBa$_2$Cu$_3$O$_{6.95}$ and CeCoIn$_5$. Considering all symmetry-lowering possibilities of the point group of interest, $C_{4v}$, we find our results to be remarkably consistent with YBa$_2$Cu$_3$O$_{6.95}$ being near vertical nematic QCP, and CeCoIn$_5$ being near diagonal nematic QCP. Our results motivate search for diagonal nematic fluctuations in CeCoIn$_5$.

\end{abstract}

\date{\today \ [file: \jobname]}

 \maketitle

\section{Introduction}

The experimental study of quantum criticality associated with a quantum critical point 
(QCP) located inside a superconducting phase is challenging due to the dominance of superconducting response.
Traditional probes for detecting effects of quantum critical fluctuations, such as transport and specific heat, are frequently overwhelmed by superconductivity. For this reason, recent experiments observing anomalous behavior in the temperature dependence of the penetration depth (PD), which is a direct measure of the superfluid density, near putative QCP's have raised hope in using this fundamental observable for a superconductor (SC) to probe the effects of quantum critical fluctuations in a superconducting system. In particular several different nodal superconductors including heavy fermions [\onlinecite{Fisher02, Chia03, Ozcan03, Shibauchi13, Truncik13}], iron-pnictides [\onlinecite{Shibauchi13}] and organic superconductors [\onlinecite{Carrington99}] have shown super-linear $T$-dependence of low temperature PD (as opposed to $T$-linear dependence expected of nodal superconductors [\onlinecite{Hardy93}]).
%The temperature dependence of PD has been shown to obey anomalous power law scaling in several different families of materials, including heavy fermions [\onlinecite{Fisher02, Chia03, Ozcan03, Shibauchi13, Truncik13}], iron-pnictides [\onlinecite{Shibauchi13}] and organic superconductors [\onlinecite{Carrington99}], all of which are in close proximity to QCPs. 
This unusual temperature dependence in a narrow doping range invites one to invoke quantum criticality.

Proximity to a putative antiferromagnetic QCP in the systems investigated in Ref.[\onlinecite{Shibauchi13}] led Hashimoto et al. to conjecture that the antiferromagnetic quantum critical fluctuations cause the observed superlinear scaling. However, since the antiferromagnetic order parameter field carries a finite momentum that does not nest the nodes, it cannot couple to nodal quasiparticles linearly while preserving crystal momentum. Hence the coupling between nodal quasiparticles and antiferromagnetic quantum critical fluctuations are irrelevant in the renormalization group sense and unlikely to alter the temperature dependence of PD in qualitative manner. Therefore to test the possibility of the quantum critical fluctuation driven anomalous scaling scenario, the search for the candidate quantum critical fluctuations should be broadened. 

%However, this opens another conundrum: low temperature PD appears nominally $T$-linear in cuprates near optimal doping [\onlinecite{Hardy93}], despite growing evidence of quantum critical fluctuations [\onlinecite{Taillefer09, Sebastian14}]. 

\begin{figure}[t]
\begin{centering}
\includegraphics[width=0.6\linewidth]{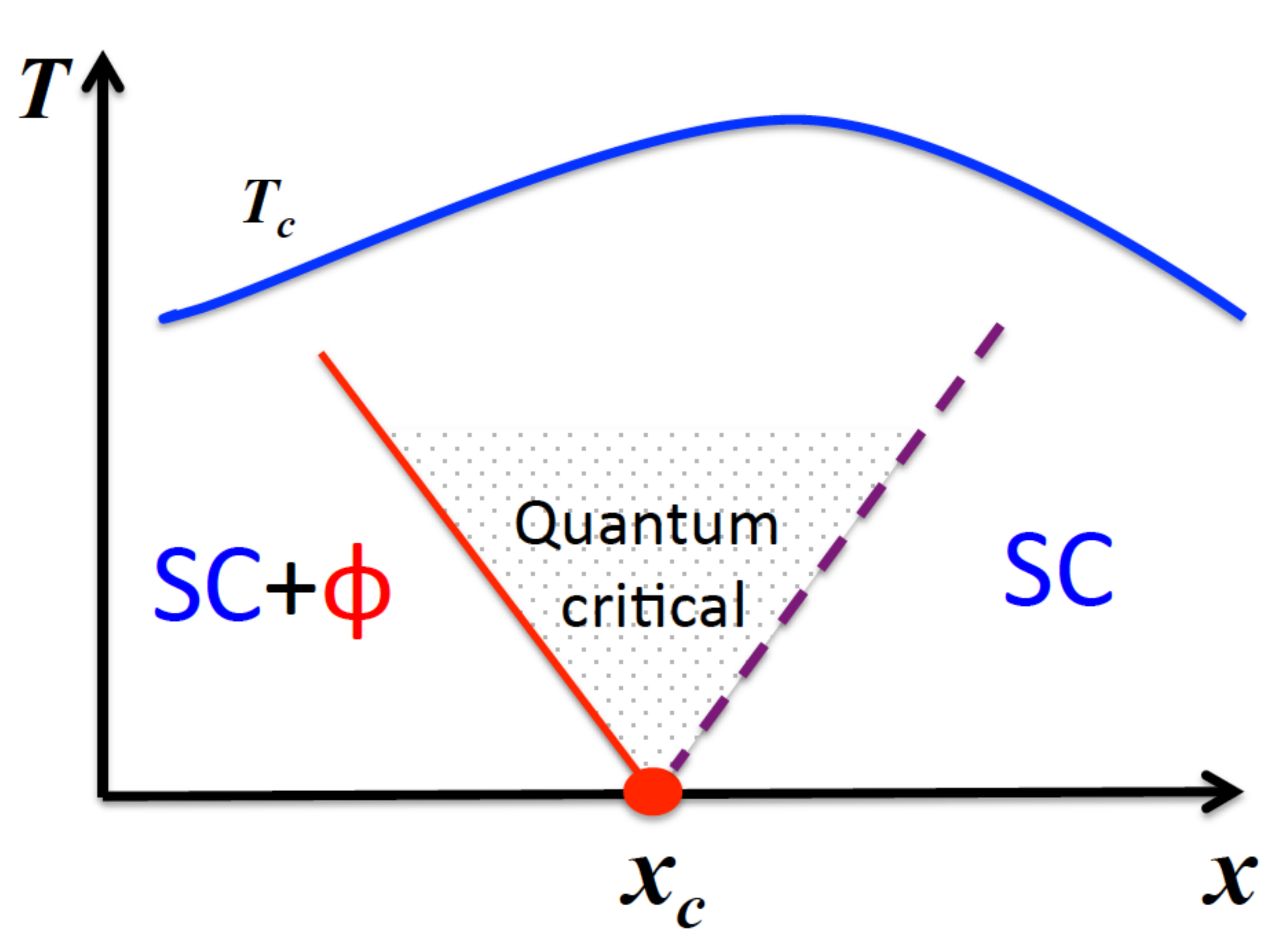} 
\end{centering}
\caption{(Color online) Schematic phase diagram of a quantum phase transition inside the superconducting dome associated with developing a new order represented by an order parameter $\phi$. Here $T_c$ is the superconducting transition temperature, and $x$ is a tuning parameter that drives the system through the quantum critical point at $x_c$. Near $x_c$ there is a quantum critical region where the penetration depth displays anomalous scaling. }
\label{Phase}
\end{figure}

\begin{figure}[t]
\begin{centering}
\subfigure[]{
\includegraphics[width=0.35\linewidth]{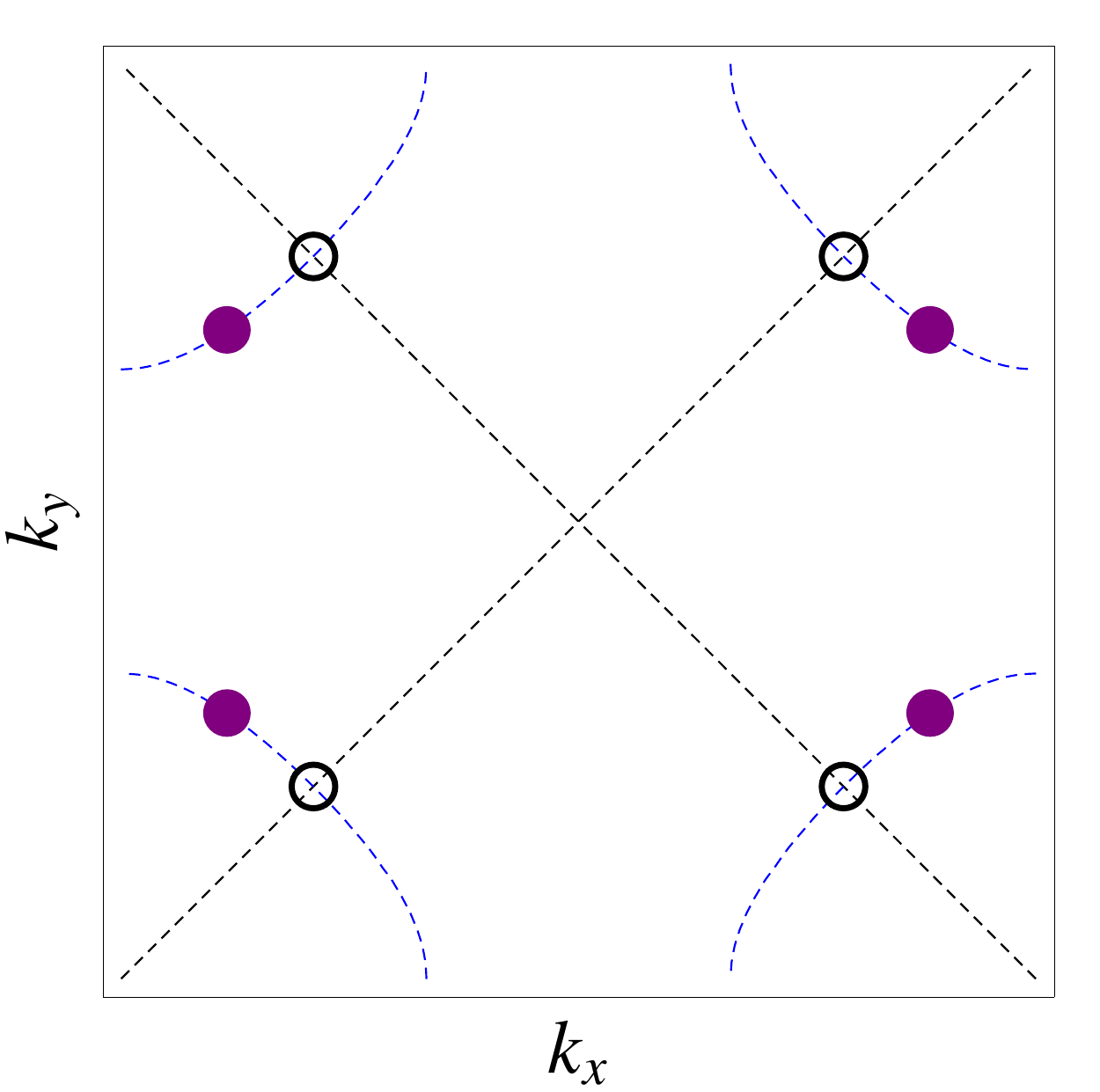} ~~}
\subfigure[]{
\includegraphics[width=0.35\linewidth]{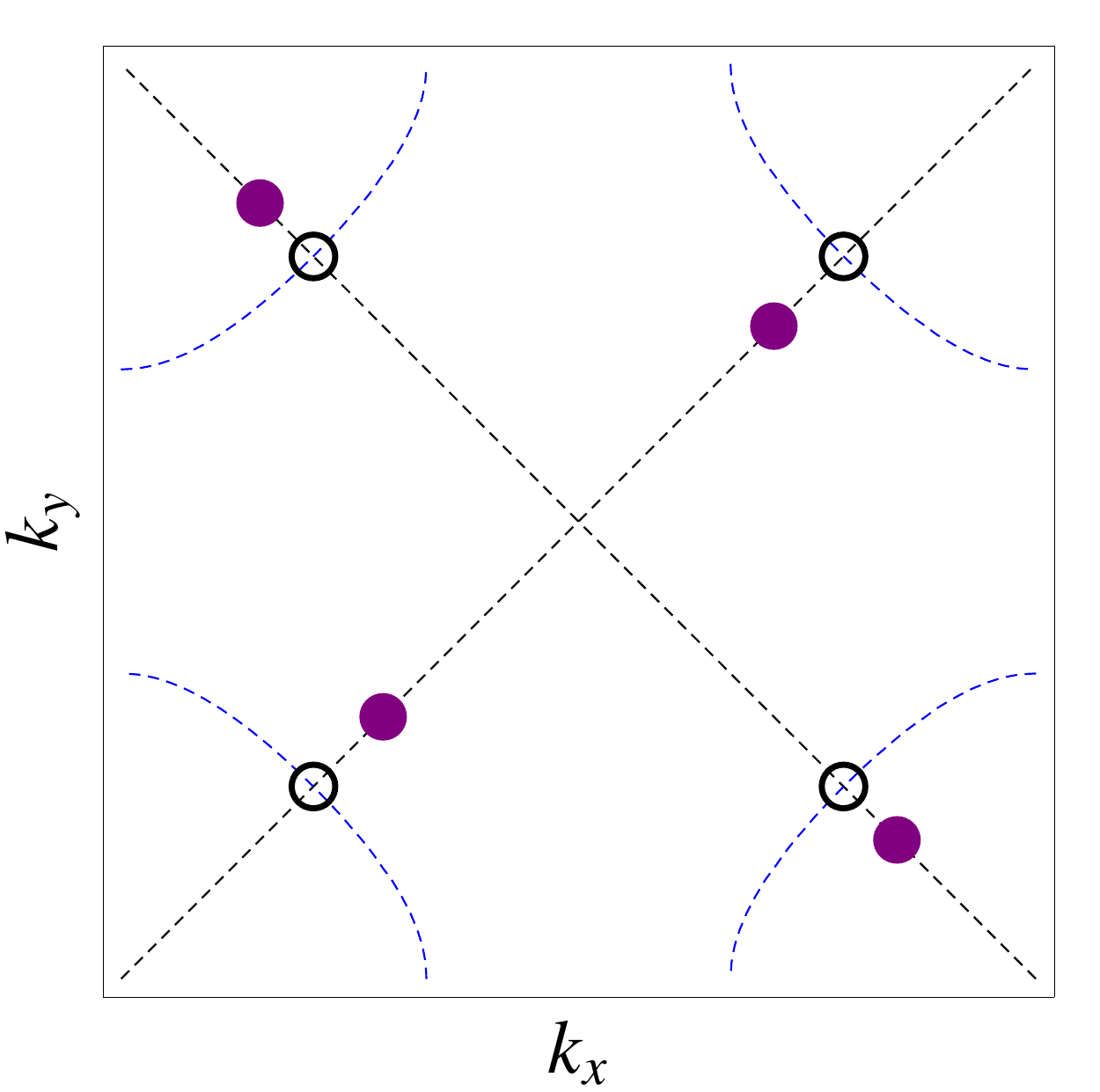} ~~}
\subfigure[]{
\includegraphics[width=0.35\linewidth]{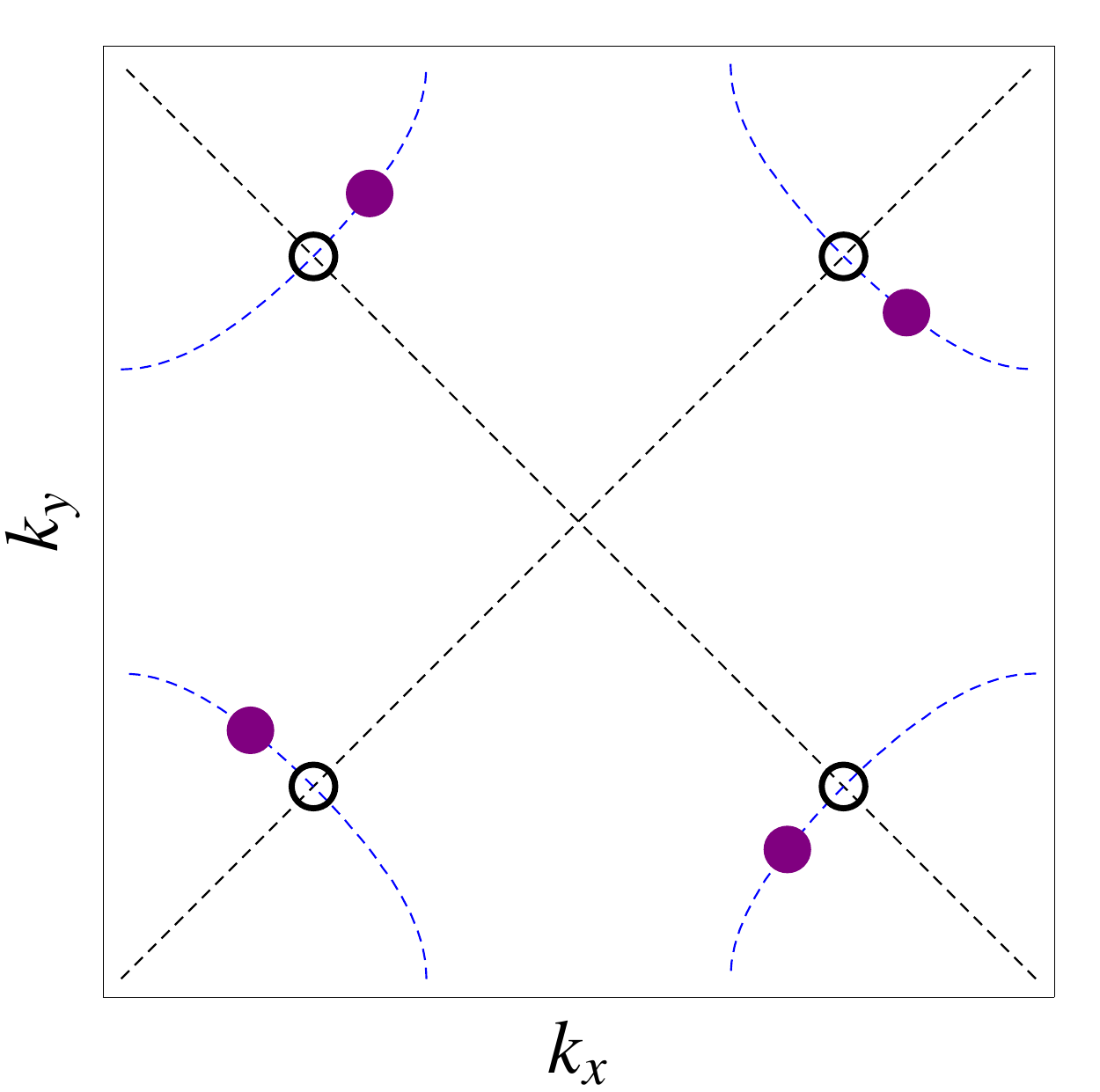} ~~}
\subfigure[]{
\includegraphics[width=0.35\linewidth]{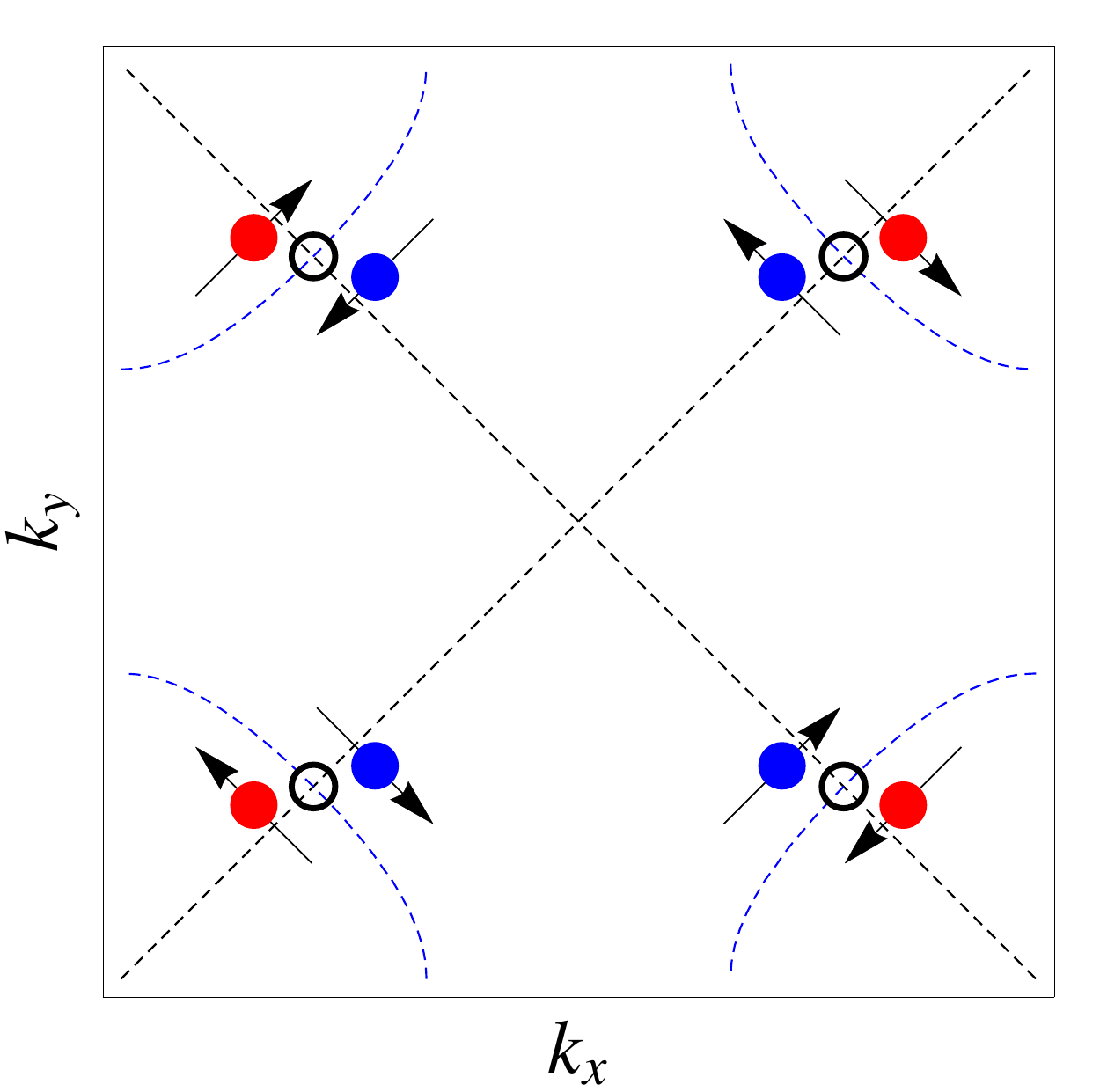} }
\end{centering}
\caption{(Color online) Nodal structure of $d_{x^2-y^2}$-SC with coexisting orders for a single band system. For multi-band systems like CeCoIn$_5$, the nodes on different surfaces will move simultaneously. The dashed (blue) lines at the corner of the Brillouin zone represent the Fermi surface, and the diagonal dashed (black) lines represent where the gap vanishes.
The open circles denote the original nodes of $d_{x^2-y^2}$-SC, and the filled circles denote the reconstructed nodes with coexisting orders. (a) B$_1$ axial nematic case; (b) B$_2$ diagonal nematic case; (c) A$_2$ case; (d) E case.  Note for (a-c), each nodal point is spin degenerate. (a, c) correspond to deformation of the gap (particle-particle channel), and hence are associated with $\tau_x$ in Nambu space. (b) corresponds to deformation of the Fermi surface (particle-hole channel), and hence $\tau_z$. The spin degeneracy is lifted 
for (d) and the blue and red filled circles have different spin-momentum helicity. The arrows in (d) indicate in-plane spin orientation at the nodal points, and here the ordering is predominantly in the particle-hole channel.  }
\label{nodes}
\end{figure}

In this paper, we explore other avenues for  a QCP inside the superconducting dome (see Fig.~\ref{Phase}) altering the temperature dependence of the PD of nodal superconductors. In particular, we will be interested in QPT's that preserve the nodes [\onlinecite{Sato06, BergChen08}]. This narrows the possibilities to ${\bf Q}=0$ time-reversal symmetric point group lowering transitions including different types of nematic order [\onlinecite{Fradkin10}]. 
These possibilities which consist of QPT's associated with one-dimensional (Fig.~\ref{nodes}(a-c)) and two-dimensional representations (see Fig.~\ref{nodes}(d)) of the point group and they
exhaust all cases of quantum critical fluctuations that can couple linearly with the nodal quasi-particles through non-derivative coupling. 

For the QCP's associated with the one-dimensional representations(Fig.~\ref{nodes}(a-c)), we compute  leading temperature dependence of the PD and
find logarithmic corrections to the $T$-linear PD expected of nodal superconductors away from QCP's. 
For the QCP associated with the  two-dimensional representation  (see Fig.~\ref{nodes}(d)) we only 
state the form of the action for completeness, without carrying out the full calculation of PD. This decision is based on two reasons: 1) the calculation of temperature dependent PD for such QCP requires an involved calculation due to having two
coupling terms that compete with each other, 2) such doubling of nodes, should it happen, will be evidenced more directly via other observables. We then compare our results to the PD measurements of YBa$_2$Cu$_3$O$_{6.95}$ and CeCoIn$_5$.

\section{The Model}
Although the nodal superconductors studied in Ref.~[\onlinecite{Fisher02, Chia03, Ozcan03, Shibauchi13, Truncik13, Carrington99}] all differ in terms of details of their nodal structure and the origin of nodes, we will focus on models of point nodes in two-dimensional system as a representative example. Furthermore, for concreteness, we consider the nodal superconductor to be a $d_{x^2-y^2}$-SC on a two-dimensional square lattice with appropriate number of nodal Dirac fermion species: two per spin for cuprates, six per spin for CeCoIn$_5$. Below we first present the single band model with two species of Dirac fermions, and then extend it to the three band model relevant for CeCoIn$_5$.

\subsection{Single band model: the case of cuprates}

Targeting at cuprates, the QCP associated with the quantum phase transition inside the dome of $d_{x^2-y^2}$-SC has been studied extensively using a single band model emplying field theoretical approaches [\onlinecite{Vojta00, Kim08, Huh08,XuQi08, Fritz09, Kleinert11, Wang12, Wang13}]. In particular an ``infinite anisotropic" fixed point was discovered in Ref.~[\onlinecite{Kim08}] associated with nodal nematic QCP, and a full renormalization group (RG) theory for the fixed point was developed in Ref.~[\onlinecite{Huh08}]. Following these references, the effective action for the nodal quasi-particles of this $d$-wave superconductor is obtained by linearizing the Bogoliubov-de Gennes mean field Hamiltonian near the two nodes ${\bm K}_1=(K, K)$, ${\bm K}_2=(-K, K)$ and their time-reversal partners. We define the momentum deviation from the nodal points ${\bm p}={\bm k}-{\bm K}_A$ near each node $A=1,2$, and the nodal Dirac fermions in
terms of the Nambu spinors $\Psi_{A{\alpha}}({\bm p})=( c_{{\bm K}_A+{\bm p}, \alpha},~ \epsilon_{\alpha\beta}c^{\dagger}_{-({\bm K}_A+{\bm k}), \beta})^T$, where $\epsilon_{\alpha\beta}$ is the antisymmetric Levi-Civita symbol, and $\alpha, \beta$ are spin indices. The effective action is
\begin{eqnarray}
S_{\Psi}&=&\sum_{{\bm p}n\alpha}\Psi^{\dagger}_{1\alpha} \left( -i\omega_n+v_Fp_x\tau_z+v_\Delta p_y\tau_x \right)\Psi_{1\alpha}\nonumber\\
&+&\sum_{{\bm p}n\alpha}\Psi^{\dagger}_{2\alpha} \left( -i\omega_n+v_Fp_y\tau_z+v_\Delta p_x\tau_x \right)\Psi_{2\alpha},
\label{eq:Dirac}
\end{eqnarray}
where $\omega_n$ is the Matsubara frequency, $v_F$ and $v_\Delta$ are the Fermi velocity and gap velocity respectively, $\tau_z$ and $\tau_x$ are Pauli matrices in the Nambu space that represent kinetic energy and pairing terms respectively. The momentum ${\bm p}=(p_x, p_y)$ has been rotated by $\pi/4$ to simplify the notation. Appendix I demonstrates how this effective action gives rise to a linear-in-$T$ temperature dependence of the PD.

The complete catalogue of time-reversal symmetric  point group lowering possibilities must correspond to the non-trivial representations of the point group of interest. For C$_{4v}$, these representations are: A$_2$, B$_1$, B$_2$ and E, where the first three are one-dimensional and the E representation is two dimensional. Associated with each of these representations, we construct the corresponding order parameter fields whose number of components equals the dimensionality of the representation.  For A$_2$, B$_1$ and B$_2$, the order parameter actions are all of the form
\begin{equation}
   S_\phi = \frac{1}{2} \int d^2rd\tau \bigg[\left(\partial_\tau\phi\right)^2 + \left(\nabla \phi\right)^2 + (x-x_c) \phi^2 + \cdots\bigg]
\label{Eq:Sphi}
\end{equation}
though they each couple differently to the quasi-particles. For the E representation case, the order parameter action is similar but with $\phi\to\phi_a$ for $a=x,y$.

To obtain an intuitive picture of the four types of phase transitions, we plot their effects on the nodes in Fig.~\ref{nodes}. For simplicity we are showing the case of single Fermi surface, and it can be easily extended to multiple Fermi surfaces. As can be seen in Fig.~\ref{nodes}(a) and (b), the B$_1$ and B$_2$ cases are axial and diagonal nematic or orientational orderings respectively. The A$_2$ case rotates the nodes clockwise or counter-clockwise. Finally, for the E case which breaks inversion symmetry, time reversal symmetry requires a Rashba type spin-orbit coupling that splits the nodes. 
Mean-field order parameters for shifting nodes along the tangent or normal of the Fermi surface involve bilinears of  Nambu spinors combined with $\tau_x$ or $\tau_z$ respectively. From this perspective, B$_1$(axial nematic) and A$_2$ channels involve $\tau_x$, B$_2$ (diagonal nematic) channel involves $\tau_z$, and the E channel involves both.

Now we turn to the leading couplings between the order parameter fields and the nodal quasiparticles. This coupling is generally of the form
\begin{equation}
S_{\Psi\phi}=\int d^2rd\tau \phi\left(\Psi^{\dagger}_1\Gamma_1\Psi_1
+\Psi^{\dagger}_2\Gamma_2\Psi_2 \right),
\label{interaction}
\end{equation}
where we rescaled the bosonic field by the coupling strength, $\phi\to \phi/g$, and used the four-component Nambu spinor $\Psi=(\Psi_{\uparrow}, \Psi_{\downarrow})^T$.  For the scalar field cases, the couplings are diagonal in spin space with $\Gamma_{1}= -\Gamma_{2} = \tau_x$ for the A$_2$ case, $\Gamma_{1} = \Gamma_2 = \tau_x$ for the B$_1$  case and $\Gamma_1 = \Gamma_2 = \tau_z$ for the B$_2$ case.  We will see that the important distinction between the three cases here is not the signs of the couplings but whether they involve $\tau_x$-type coupling that belong to the pairing channel and $\tau_z$-type coupling that belong in the particle-hole channel. For the E case, where $\phi$ is a linear combination of $\phi_{x, y}$, we have $\Gamma_{1, 2}=(\sigma_x\pm\sigma_y)\tau_z+\lambda(\sigma_x\pm\sigma_y)\tau_x$, where $\sigma_{x, y}$ are the spin Pauli matrices, and $\lambda$ denotes the ratio between the couplings in particle-particle and particle-hole channels. With two competing couplings, the calculation is more involved for the E case and we will leave it for a future study.

\subsection{Multiband model: the case of CeCoIn$_5$}

Now we generalize the above single-band low energy effective model to multi-band systems in order to describe the temperature dependence of penetration depth for CeCoIn$_5$ at low temperatures. 
CeCoIn$_5$ is electronically quasi two-dimensional [\onlinecite{Petrovic01, Sarrao01, Thompson02}].  Its superconducting gap has point-like nodes, and most likely has $d_{x^2-y^2}$ symmetry, with point nodes in two dimensions along ${\bm k}=(0, 0)\to(\pm 1, \pm 1)\pi/a$ directions, as shown by the magnetic field-angle dependence of thermal conductivity [\onlinecite{Izawa01}] and specific heat [\onlinecite{An10}], and more recently by STM quasiparticle interference spectroscopy [\onlinecite{Allan13, Zhou13}]. Hence the low energy degrees of freedom in the superconducting state of CeCoIn$_5$ can be modeled by two dimensional Dirac fermions around the nodal points. The electronic structure has $C_{4v}$ symmetry, and hence the corresponding new phases can be classified according to the irreducible representations of $C_{4v}$ as shown above for cuprates. 

However, there is one extra complication for CeCoIn$_5$ as compared to cuprates: there are three sets of nodal points associated with the three different Fermi surfaces \footnote{Its Fermi surface consists of a small hole-like Fermi surface ($\beta$) centered at the $\Gamma$ point, and two electron-like Fermi surfaces ($\alpha_1$, $\alpha_2$) centered at $(\pi, \pi)$ [\onlinecite{Allan13, Zhou13}]. The dominant superconducting gap is on the $\alpha_1$ Fermi surface, and there is smaller gap at the $\alpha_2$ and $\beta$ Fermi surfaces [\onlinecite{Allan13, Zhou13}].}. Hence the low energy degrees of freedom in the superconducting state involve three sets of nodal Dirac fermions. The fermion action becomes
\begin{eqnarray}
S_{\Psi}&=&\sum_{{\bm p}na\alpha}\Psi^{\dagger}_{1a\alpha} \left( -i\omega_n+v_F^ap_x\tau_z+v_\Delta^a p_y\tau_x \right)\Psi_{1a\alpha}
\nonumber\\ &+&\sum_{{\bm p}na\alpha}\Psi^{\dagger}_{2a\alpha} \left( -i\omega_n+v_F^ap_y\tau_z+v_\Delta^a p_x\tau_x \right)\Psi_{2a\alpha},
\end{eqnarray}
with the three sets of velocities $(v_{F a}, v_{\Delta a})$, where $a=1, 2, 3$, labeling different Fermi surfaces. The Dirac fermions couple to the order parameter fluctuations with different coupling strength $g_a$, with the action
\begin{equation}
S_{\Psi\phi}=\sum_a\int d^2rd\tau g_a \phi\left(\Psi^{\dagger}_{1a}\Gamma_{1a}\Psi_{1a}
+\Psi^{\dagger}_{2a}\Gamma_{2a}\Psi_{2a} \right).
\end{equation}
The bare action of the order parameter field remains the same as in the single band model (Eq.~(\ref{Eq:Sphi})).

\begin{table}[t]
    \begin{tabular}{ | c | c | c | c |c | c | c |c|}
    \hline
    order parameter& coupling & $\left(\frac{v_\Delta}{v_F}\right)^*$ & $\delta\lambda^{-2}(T)$ \\ \hline
 B$_1$, A$_2$ &    $\tau_x$& $0$  & $\frac{T}{T_0}\left( 1+\frac{1}{l_0}\ln\frac{T_0}{T} \right)$ \\ \hline
B$_2$ & $\tau_z$ &$\infty$ & $\frac{T}{T_0}\left(1+\frac{1}{l_0}\ln\frac{T_0}{T}\right)^{-1}$ \\ \hline
    \end{tabular}\par
\caption{Summary of main results: the representation of the new order parameter, the involved Nambu space Pauli matrix for the coupling, fixed point value of the velocity ratio under RG, and the temperature dependence of penetration depth. Here $T_0$ is the 
temperature 
corresponding to the RG scale $l_0$, which is related to the initial value of the velocity ratio by $\left(\frac{v_\Delta}{v_F}\right)_0\simeq\frac{\pi^2N}{8}\frac{1}{l_0\ln l_0}$ for $\tau_x$-coupling,
and $\left(\frac{v_F}{v_\Delta}\right)_0\simeq\frac{\pi^2N}{8}\frac{1}{l_0\ln l_0}$ for $\tau_z$-coupling.} 
\end{table}

\section{$T$-dependence of the Penetration Depth} 

 The $T$-dependence of PD near the putative QCPs can be determined by the RG equation for the paramagnetic part of the electromagnetic response kernel $K(T)$, with $\lambda^{-2}(T)=4\pi K(T)$. 
In order to derive the necessary RG equations, we use the standard $1/N$ expansion which is known to be controlled by the smallness of the velocity ratio (e.g., $v_\Delta/v_F$) near the infinitely anisotropic fixed points even for realistic value of $N=2$ [\onlinecite{Huh08}].

\subsection{Single band model}

\begin{figure}
\begin{centering}
\subfigure[]{
\includegraphics[width=0.6\linewidth]{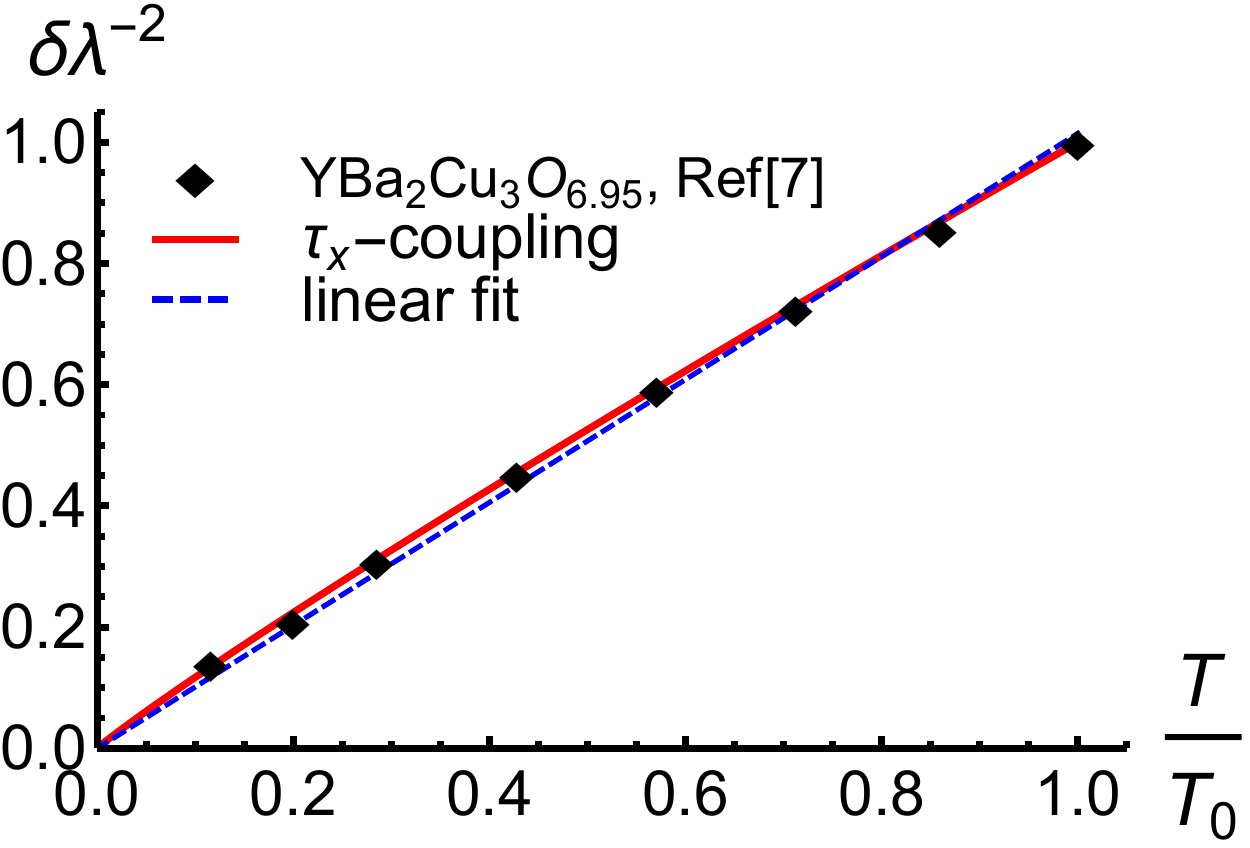} 
}
\subfigure[]{
\includegraphics[width=0.6\linewidth]{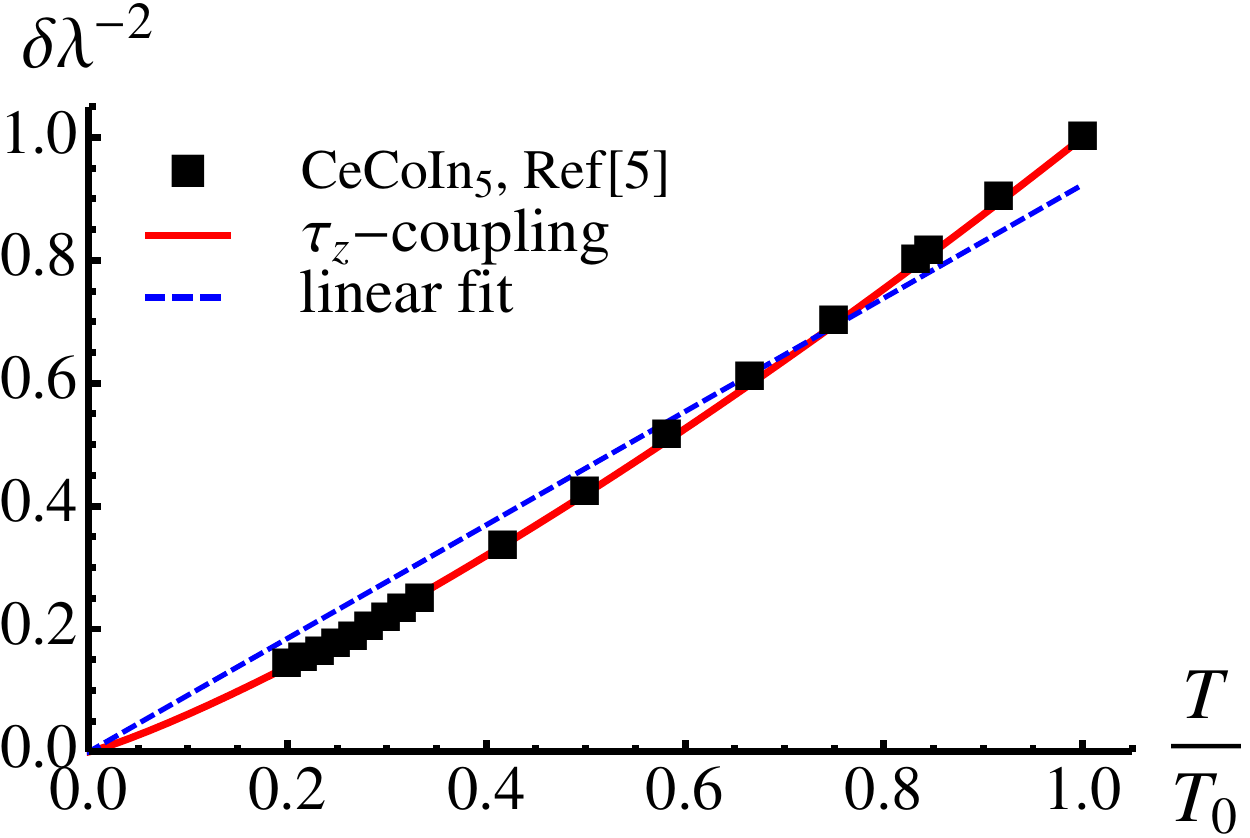} 
}
\end{centering}
\caption{(Color online) Temperature dependence of the penetration depth (or equivalently the superfluid density with $\delta\rho_s\sim \delta\lambda^{-2}$), fitting experimental results for YBa$_2$Cu$_3$O$_{6.95}$ (a) and CeCoIn$_5$ (b). The diamonds and squares are the experimental results for YBa$_2$Cu$_3$O$_{6.95}$ from [\onlinecite{Hardy93}] and for CeCoIn$_5$ from [\onlinecite{Truncik13}] respectively. The experimental values of $\delta\lambda^{-2}$ have been normalized by an overall factor so that $\delta\lambda^{-2}(T_0)=1$. The solid lines are fit to the analytic expressions shown in Table I. The fitting parameters are: (a) for $\tau_x$-coupling, $1/l_0=0.075$ as deduced from the experimentally measured velocity anisotropy $v_F/v_\Delta\sim 14$  [\onlinecite{Chiao00, Norman95}] with $N=2$; (b) for $\tau_z$-coupling with $N=6$, $1/l_0=0.2756$ as best fit, and correspondingly $v_F^a/v_\Delta^a\simeq 1.5829$ for all three bands $a=1, 2, 3$.}
\label{scaling}
\end{figure}

For the single band model, the response kernel $K(T)$ satisfies the following simple homogeneity relation (see Appendix II):
 \begin{equation}
K(\Lambda_0, v_F^0, v_\Delta^0)=e^lK(\Lambda, v_F(l), v_\Delta(l)),
\label{Eq.scaling}
\end{equation}
with the energy scale $\Lambda=\Lambda_0 e^{-l}$, where $\Lambda_0$ is the UV energy scale, $v_F^0, v_\Delta^0$ are the velocities at scale $\Lambda_0$, and $v_F(l), v_\Delta(l)$ are the corresponding velocities at scale $\Lambda$.
Eq.~\eqref{scaling} states that the scaling of $K$ is determined by scaling of the two velocities $v_F$ and $v_\Delta$. This is a result of the cancellation of the renormalization of the fermion field and fermion-electromagnetic field coupling which prevents renormalization of the coupling between the fermion current and the electromagnetic gauge field. Hence the anomalous dimension for the current-current correlation function vanishes. 
But by dimensional analysis, $K(\Lambda, v_F, v_\Delta)=\Lambda f(v_F/v_\Delta)$, with a scaling function $f$. Then we obtain the RG equation for $K$ using the scaling relation Eq.(\ref{Eq.scaling}) to be:
\begin{equation}
\frac{d\ln K}{d\l}= -1+ \frac{d\ln (v_F/v_\Delta)}{dl}.
\label{RG}
\end{equation} 
This shows that the RG flow of the kernel $K$ is entirely determined by the flow of the velocity anisotropy ratio $v_F/v_\Delta$.

The RG flow of the velocity anisotropy ratio $v_\Delta/v_F$
was calculated for $\tau_x$-coupling (e.g., B$_1$ and A$_2$ channel ordering, see Fig\ref{nodes}(a), (c)) in [\onlinecite{Huh08}], and the corresponding RG flow for $\tau_z$-coupling (e.g., B$_2$ channel ordering, see Fig\ref{nodes}(b)) can be obtained by a ``duality" transformation (exchanging $v_\Delta$ with $v_F$, and exchanging the functional dependence of their RG coefficients, see supplementary material S2). The fixed points associated with each of these flows are  $(v_\Delta/v_F)^*=0$ for $\tau_x$-coupling, and $(v_F/v_\Delta)^*=0$ for $\tau_z$-coupling. 
Using these RG flows we now derive the asymptotic temperature dependence of $K(T)$ near the fixed points. Consider first $\tau_x$-coupling, where the velocity anisotropy ratio is of the asymptotic form $v_\Delta/v_F\simeq (\pi^2N/8)(1/l\ln l)$ [\onlinecite{Huh08}], and hence 
$\frac{d\ln (v_\Delta/v_F)}{dl}\simeq -1/l$. Carrying out a similar process also for $\tau_z$-coupling one obtains 
\begin{equation}
K\sim \left\{ 
\begin{array}{lcl} le^{-l}& \mbox{for} & \tau_x \mbox{-coupling}, \\  l^{-1}e^{-l}& \mbox{for} & \tau_z \mbox{-coupling},
\end{array}\right.
\end{equation}
where the exponential decay comes from the engineering dimension of $K$, and the power law prefactor arising from the coupling of nodal fermions to critical modes. 
Using the RG equation for temperature, $d\ln T/dl\simeq -1$, one then obtains the asymptotic temperature dependence of the kernel near fixed points as shown in Table I. The only free parameter here is the initial value of the velocity ratio.

\subsection{Multiband model}

The above calculational procedure can be straightforwardly generalized to the multiband models.
In the three band model for CeCoIn$_5$, there are three coupling constants $g_1$, $g_2$, $g_3$ that flow. However, it turns out they flow to a stable fixed point where $g_1^*=g_2^*=g_3^*$ exhibiting emergent enlarged symmetry (see Appendix II). Therefore 
the behavior of the system near the fixed point is well captured by a single band model with 
$v_F/v_\Delta$ in the  RG equation ({\ref{RG}}) replaced by 
a summation of the velocity ratios from different bands:
\begin{equation}
\frac{d\ln K}{dl}=-1+\frac{d\ln R_v}{dl},
\end{equation}
where $R_v=\sum_a(v_{F a}/v_{\Delta a})$.

The asymptotic form of the velocity ratios remains of the same logarithmic form as in the single band model, since the logarithm comes from integration over the fermion propagator, and the asymptotic form of the fermion propagator does not change with the presence of multiple bands.  As a result, the RG equation of the electromagnetic kernel, and consequently the temperature dependence of the penetration depth, are of the same asymptotic form as in the single band model.

\subsection{Comparison with experiments}

 Our result shows that coupling between nodal quasiparticles and quantum critical fluctuations cause logarithmic corrections to the $T$-linear temperature dependence of a nodal superconductor's inverse penetration depth rather than a universal power law. Such logarithmic corrections would yield ``apparent'' power law with
sublinear (superlinear) $T$-dependence of the PD depending on coupling in the particle-particle $\tau_x$ channel (particle-hole $\tau_z$ channel). 
The ``apparent'' exponent will be non-universal and depend on the bare velocity anisotropy ratio.

Now we use the predicted $T$-dependence of the PD (Table I) to fit experimental results for YBa$_2$Cu$_3$O$_{6.95}$ (optimal doping) and for CeCoIn$_5$.
The temperature dependence of $\delta\lambda^{-2}$ for YBa$_2$Cu$_3$O$_{6.95}$ has been favorably compared with $T$-linear behavior expected for $d$-wave superconductor with nodal quasi-particles, despite growing evidence of quantum critical fluctuations [\onlinecite{Taillefer09, Sebastian14}]. However fitting the data from Ref.~[\onlinecite{Hardy93}] with $\tau_x$-coupling (e.g. axial nematic order, B1 case in Fig.~\ref{nodes}a), we find the extreme value of velocity anisotropy  $v_F/v_\Delta\simeq 14$ [\onlinecite{Chiao00, Norman95}]
implies that the existence of nematic quantum critical fluctuations in YBa$_2$Cu$_3$O$_{6.95}$ cannot be ruled out. 

The experimental data in [\onlinecite{Truncik13}] for CeCoIn$_5$ are best fit by $\tau_z$-coupling (e.g., diagonal nematic order, B2 case in Fig.~\ref{nodes}b) with moderate velocity anisotropy ratio of $v_F^a/v_\Delta^a\simeq 1.5829$ for all three bands (see Fig.~\ref{scaling}b)\footnote{The particular value of $(v_F/v_\Delta)_0$ depends on higher order corrections (say in $1/l$). Generally the apparent power increases with increasing $(v_F/v_\Delta)_0$.}. Thus the observation of superlinear $T$-dependence of PD in CeCoIn$_5$ is consistent with logarithmic correction to penetration depth due to quantum critical diagonal nematic fluctuations.

\section{Conclusion}

In conclusion, we have shown in this paper that inside a two dimensional nodal superconductor with point nodes, quantum critical fluctuations associated with TRS-preserving ${\bm Q}=0$ orders that have nonvanishing amplitude at the nodal points give rise to logarithmic corrections to the temperature dependence of PD (see Table I). In particular, we find the direction of fluctuations in the nodal positions play the key role: fluctuations along the direction normal (tangent) to the underlying Fermi surface cause apparent super- (sub-) linear scaling of the PD as a function of temperature. Our results are qualitatively different from predictions of $T$-linear to $T$-square crossover behavior in PD of different origins, e.g. impurity scattering [\onlinecite{Hirschfeld93}], nonlocal effects [\onlinecite{Leggett97}], antiferromagnetic fluctuations [\onlinecite{Nomoto13}]. 

With its normal state displaying non-Fermi liquid behavior, CeCoIn$_5$ has long been suggested to be close to a QCP [\onlinecite{Petrovic01, Sarrao01, Thompson02, Bianchi03}]. The QCP is associated with a magnetic phase transition based on empirical evidence: for example, replacing 15-30$\%$ In by Cd produces an antiferromagnetic ground state [\onlinecite{Pham06}]. Our analysis of the temperature dependence of PD, and also the studies on zero-temperature PD in [\onlinecite{Sachdev13, Levchenko13}], reveal that the magnetic QCP cannot be the whole story. In fact, 
our result suggests that CeCoIn$_5$ may be close to a QCP of diagonal nematic order. 

Interestingly, previous theoretical studies 
on quantum critical diagonal nematic fluctuations [\onlinecite{Doh06, Kao07}] have shown that
continuous quantum phase transition associated with Pomeranchuk instability [\onlinecite{Oganesyan01}] is possible with the diagonal nematic even in the presence of lattice unlike the axial nematic case. Further they also showed that such fluctuations cause non-Fermi liquid behavior. 

A direct confirmation of strong nematic fluctuations in 
 CeCoIn$_5$  would enlarge the class of materials exhibiting point group symmetry breaking. This has so far been established in the cuprates [\onlinecite{Lawler10, Mesaros11}], iron-pnictides [\onlinecite{Chu10, Chuang10}], and URu$_2$Si$_2$ [\onlinecite{Okazaki11, Ikeda12}]. Furthermore, establishment of quantum criticality inside superconducting phase would contribute to the
 emerging picture of intertwined orders in strongly correlated electron systems [\onlinecite{Fradkin14, Davis13}]. 
Hence detection of nematic fluctuations in CeCoIn$_5$ can offer valuable experimental and theoretical insights. This can be achieved following
recent developments in using response to strain to measure nematic fluctuations [\onlinecite{Chu12, Meingast14, Riggs14}].

We thank Steve Kivelson for helpful discussions. J.-H.S and E.-A.K were supported by the U.S. Department of Energy, Office of Basic Energy Sciences, Division of Materials Science and Engineering under Award de-sc0010313.

\onecolumngrid
\appendix

\section*{Appendix I: Penetration depth for $d$-wave superconductor}
In this section, we derive the $T$-dependence of penetration depth for a $d$-wave superconductor from the action of free Dirac fermions (Eq.(1) in main text). The electromagnetic kernel at zero external frequency can be written as
\begin{equation}
K_{ij}({\bm q}, T)= T\sum_{nA}\int\frac{d^2{\bm p}}{(2\pi)^2}v_{\bm k}^iv_{\bm k}^j {\rm Tr} \left[G^{(0)}_A({\bm p}_+,\omega_n)G^{(0)}_A({\bm p}_-,\omega_n)\right],
\end{equation}
with the fermion Green's functions near the two pairs of nodes $[G^{(0)}_1]^{-1}=-i\omega_n+v_Fp_x\tau_z+v_\Delta p_y\tau_x $ and $[G^{(0)}_2]^{-1}=-i\omega_n+v_Fp_y\tau_z+v_\Delta p_x\tau_x $. Here to compare with the result of Kosztin and Leggett [\onlinecite{Leggett97}], we consider also $K$ at finite momenta. The Fermi velocity $v_{\bm k}^i$ can be approximated by its value at the corresponding nodal point. The momentum summation is only non-zero when $i=j$, and $v_{\bm k}^i v_{\bm k}^i=(v_{{\bm K}_A}^i)^2=v_F^2/2$. After taking the trace, integrating over momentum, replacing the frequency summation by an integral, and defining $x=\omega/T$, one obtains
\begin{equation}
K({\bm q}, T)\sim \frac{v_F}{v_\Delta}T\int dx\frac{1}{e^x+1}\left[ 1-\sqrt{1-\left( \frac{2Tx}{v_Fq} \right)^2}\right].
\end{equation}
We note that the above result is of the scaling form 
\begin{equation}
K(q, T)\sim T^{\alpha_K/z} {\cal F}(v_Fq/T^{1/z}),
\end{equation}
 with exponent $\alpha_K=1$ and dynamical exponent $z=1$. 

For a specular boundary, the temperature dependent part of penetration depth can be determined from 
\begin{equation}
\frac{\delta\lambda_{\rm spec}(T)}{\lambda_0}=-\frac{2}{\pi}\int_0^{\infty}d{\tilde q}\frac{\delta{\tilde K({\tilde q},T)}}{({\tilde q}^2+1)^2},
\end{equation}
 with ${\tilde q}=q\lambda_0$ [\onlinecite{Leggett97}]. One thus obtains for a $d$-wave superconductor
\begin{equation}
\frac{\delta\lambda_{\rm spec}(T)}{\lambda_0}= \left\{ 
\begin{array}{lcl} A T& \mbox{for} & T\gg T^*,  \\   BT^2& \mbox{for} &  T\ll T^*,
\end{array}\right.
\end{equation}
with the crossover temperature $T^*=v_F/\lambda_0\simeq(\xi_0/\lambda_0)\Delta_0$. The penetration depth crosses over from $T$-linear at higher temperatures to $T^2$ at lower temperatures, as was shown by Kosztin and Leggett in [\onlinecite{Leggett97}].

\section*{Appendix II: Renormalization of the electromagnetic kernel}

We present here the renormalization group calculation of the electromagnetic kernel for the system of nodal Dirac fermions coupled with critical fluctuations. A RG theory of such systems has been developed by Huh and Sachdev in [\onlinecite{Huh08}] for single band models. Here we extend their approach to multiband systems, and to the calculation of correlation functions. The RG coefficients for the renormalization of the fermion Green's function have been obtained for $\tau_x$-coupling in [\onlinecite{Huh08}]. The results for $\tau_z$-coupling can be obtained in a similar way, which turn out to be related to the $\tau_x$ case by a ``duality'' transformation as shown below. The RG coefficient of the current vertex is shown to be the same as the anomalous dimension of the fermion field, which ensures that the coupling between the fermion current and the electromagnetic gauge field is not renormalized.

The electromagnetic kernel is defined as
\begin{equation}
K(\Lambda, v_F, v_\Delta)\equiv\int_\Lambda {\cal D}\Psi{\cal D}\Psi^\dagger{\cal D}\phi {\bm J}({\bm 0}, 0)\cdot {\bm J}({\bm 0}, 0) e^{-S[\Psi, \phi]},
\end{equation}
where the current is
\begin{equation}
 {\bm J}({\bm q}, \Omega)=\int_\Lambda \frac{d^2{\bm k}}{(2\pi)^2} \frac{d\omega}{2\pi} {\bm v}_{\bm k}\Psi^\dagger({\bm k}+{\bm q}/2, \omega+\Omega/2)\Psi({\bm k}-{\bm q}/2, \omega-\Omega/2).
\end{equation}
To simplify the notation, we keep the band index $a=1, 2, \cdots, M$ implicit (e.g. ${\cal D}\Psi{\cal D}\Psi^\dagger\equiv \prod_{a=1}^M {\cal D}\Psi_a{\cal D}\Psi^\dagger_a$).
Adding electromagnetic field to the system introduces a coupling term ${\bm A}\cdot{\bm J}$ to the action and allows us to write $K$ as
\begin{equation}
K(\Lambda, v_F, v_\Delta)=\left.\frac{\delta^2\ln Z}{\delta A_i\delta A_i}\right\vert_{{\bm A}=0},
\label{Eq.K}
\end{equation}
with the partiton function in the presence of the electromagnetic field
\begin{equation}
Z(\Lambda)=\int_\Lambda {\cal D}\Psi{\cal D}\Psi^\dagger{\cal D}\phi  e^{-S[\Psi, \phi]-{\bm A}\cdot{\bm J}({\bm 0}, 0)}.
\end{equation}

\subsubsection*{Electromagnetic kernel of free fermions from RG}

To set the stage, we consider first free Dirac fermions coupled with the external gauge field.
The partition function reads
\begin{eqnarray}
Z(\Lambda)&=&\int {\cal D}\Psi{\cal D}\Psi^\dagger  e^{-S_{\Psi}-{\bm A}\cdot {\bm J}({\bm 0}, 0)},\\
S_{\Psi}&=& \int_0^{\Lambda} \frac{d^2{\bm k}}{(2\pi)^2} \frac{d\omega}{2\pi} \Psi^\dagger G_0^{-1}\Psi.\nonumber
\end{eqnarray}

The RG procedure we follow begins by separating slow and fast modes. For the current this implies
\begin{eqnarray}
{\bm J}({\bm 0}, 0)&=&{\bm J}_<({\bm 0}, 0)+{\bm J}_>({\bm 0}, 0),\nonumber\\
{\bm J}_<({\bm 0}, 0)&=&\int_0^{\Lambda/b} \frac{d^2{\bm k}}{(2\pi)^2} \frac{d\omega}{2\pi} {\bm v}_{\bm k}\Psi^\dagger({\bm k}, \omega)\Psi({\bm k}, \omega),\nonumber\\
{\bm J}_>({\bm 0}, 0)&=&\int_{\Lambda/b}^{\Lambda} \frac{d^2{\bm k}}{(2\pi)^2} \frac{d\omega}{2\pi} {\bm v}_{\bm k}\Psi^\dagger({\bm k}, \omega)\Psi({\bm k}, \omega).
\end{eqnarray} 

For the partition function this implies
\begin{eqnarray}
Z(\Lambda)=\int {\cal D}\Psi_<{\cal D}\Psi^\dagger_<e^{-S_\Psi^<-{\bm A}\cdot {\bm J}_<}\int {\cal D}\Psi_>{\cal D}\Psi^\dagger_> e^{-S_\Psi^>-{\bm A}\cdot {\bm J}_>}.
\label{Eq.Z0}
\end{eqnarray}
Integrating out the fast modes generates a new term in the action that is a functional of $\bm A$,
\begin{eqnarray}
e^{-\delta S_{\bm A}}=\int {\cal D}\Psi_>{\cal D}\Psi^\dagger_> e^{-S_\Psi^>-{\bm A}\cdot {\bm J}_>},
\end{eqnarray}
where the leading order term represents the superfluid stiffness 
\begin{eqnarray}
\delta S_{\bm A}\simeq -\frac{1}{2}\delta\kappa A^2,
\end{eqnarray}
with
 \begin{equation}
\delta \kappa=\int_{\Lambda/b}^\Lambda \frac{d^2{\bm k}}{(2\pi)^2} \frac{d\omega}{2\pi}\left[G_0({\bm k}, \omega)\right]^2\left({\bm v}_{\bm k}\right)^2={\cal A} \frac{v_F}{v_\Delta}\left( \Lambda-\frac{\Lambda}{b} \right).
\end{equation}
The partition function then obeys
\begin{eqnarray}
Z(\Lambda)=Z(\Lambda/b)e^{\frac{1}{2}\delta\kappa A^2}.
\end{eqnarray} 
Then one obtains from Eq.(\ref{Eq.K}),
\begin{equation}
K(\Lambda)=K(\Lambda/b)+\delta \kappa.
\end{equation}
With $b\equiv e^l$, where $l$ is infinitesimally small, one can expand the above equation to yield
\begin{equation}
\left(\Lambda\frac{d}{d\Lambda}-1\right)K(\Lambda)=0,
\end{equation}
where we have used $K_0={\cal A} \frac{v_F}{v_\Delta} \Lambda$. This then reproduces the linear $T$-dependence of the kernel.

\subsubsection*{Electromagnetic kernel for fermions coupled to critical modes from RG}

Now consider coupling the nodal fermions to the critical fluctuations. The partition function can be written as
\begin{eqnarray}
Z(\Lambda)&=&\int {\cal D}\Psi_<{\cal D}\Psi^\dagger_<{\cal D}\phi_<  e^{-S_<-{\bm A}\cdot{\bm J}_<}\int {\cal D}\Psi_>{\cal D}\Psi^\dagger_>{\cal D}\phi_> e^{-S_>-{\bm A}\cdot{\bm J}_>} e^{-S_{\rm int}}.
\end{eqnarray}
Note that since the critical fluctuations have ${\bm Q}=0$, they do not mix fermions in different bands.
The fermion-critical mode coupling now renormalizes the action of the slow modes $S_<$, and the coupling to gauge field ${\bm A}\cdot{\bm J}_<$. It also contributes to the superfluid stiffness term via coupling to $\Psi_>$. We consider these effects one by one.

Consider first the renormalization of the fermionic part of the action.
Using the bare action of slow fermions 
\begin{equation}
S^{(0)}_{\Psi}= \int_0^{\Lambda/b} \frac{d^2{\bm k}}{(2\pi)^2} \frac{d\omega}{2\pi} \Psi^\dagger_< G_0^{-1}\Psi_<,
\end{equation}
with $G_0^{-1}=-i\omega+v_Fk_x\tau_z+v_\Delta k_y\tau_x$, integrating out fast modes leads to a correction to the effective action 
\begin{equation}
S_{\Psi}=\int_0^{\Lambda/b}  \frac{d^2{\bm k}}{(2\pi)^2} \frac{d\omega}{2\pi} \Psi^\dagger_<\left[ G_0^{-1}-\delta\Sigma\right]\Psi_<,
\end{equation}
with the ``shell'' self energy
\begin{eqnarray}
\delta\Sigma({\bm k}, \omega)=\Sigma({\bm k}, \omega;\Lambda)-\Sigma({\bm k}, \omega;\Lambda/b),
\end{eqnarray}
where $\Sigma({\bm k}, \omega;\Lambda)$ is the self energy obtained by integrating modes with $0<|{\bm k}|<\Lambda$, and its cutoff dependence can be written as
\begin{eqnarray}
\frac{d}{d\ln \Lambda}\Sigma({\bm k}, \omega;\Lambda)=C_1(-i\omega)+C_2v_Fk_x\tau_z+C_3v_\Delta k_y\tau_x.
\end{eqnarray}
The action then reads
\begin{equation}
S_{\Psi}=(1-C_1l)\int_0^{\Lambda/b}  \frac{d^2{\bm k}}{(2\pi)^2} \frac{d\omega}{2\pi} \Psi^\dagger_<\left[ -i\omega+v'_F k_x\tau_z+v'_\Delta k_y\tau_x\right]\Psi_<,
\end{equation}
with the renormalized velocities for infinitesimal $l=\ln b$
\begin{eqnarray}
v'_F&=&v_F\frac{1-C_2l}{1-C_1l},\\
v'_\Delta&=&v_\Delta\frac{1-C_3l}{1-C_1l}.
\end{eqnarray} 
Then one performs the rescaling,
\begin{eqnarray}
k&=&k' e^{-l}, \\
\omega&=&\omega'e^{-l},\\ 
\Psi(k, \omega)&=&\Psi'(k', \omega')\exp\left[ \frac{1}{2}\int_0^l dl'(4-\eta_f) \right],\\
\phi(k, \omega)&=&\phi'(k', \omega')\exp\left[ \frac{1}{2}\int_0^l dl'(5-\eta_b) \right].
\end{eqnarray}
To recover the original form of the action, it is required that $\eta_f=-C_1$. The action after rescaling is then
\begin{equation}
S_{\Psi}=\int_0^{\Lambda}  \frac{d^2{\bm k}'}{(2\pi)^2} \frac{d\omega'}{2\pi} \left(\Psi'\right)^\dagger\left[ -i\omega'+v'_F k'_x\tau_z+v'_\Delta k'_y\tau_x\right]\Psi'.
\end{equation}

Consider next the renormalization of the coupling between fermions and the electromagnetic field. The bare coupling is 
\begin{equation}
S^{(0)}_{A{\Psi}}={\bm A}\cdot{\bm J}_<({\bm 0}, 0)={\bm A}\cdot \int_0^{\Lambda/b} \frac{d^2{\bm k}}{(2\pi)^2} \frac{d\omega}{2\pi}{\bm v}_{\bm k} \Psi^\dagger_< \Psi_<,
\end{equation}
which also receives corrections after integrating out fast modes, 
\begin{equation}
S_{A{\Psi}}={\bm A}\cdot \int_0^{\Lambda/b} \frac{d^2{\bm k}}{(2\pi)^2} \frac{d\omega}{2\pi}\left({\bm v}_{\bm k}+\delta{\bm\Upsilon}\right) \Psi^\dagger_< \Psi_<,
\end{equation}
with ``shell" vertex correction
\begin{eqnarray}
\delta{\bm\Upsilon}={\bm\Upsilon}(\Lambda)-{\bm\Upsilon}(\Lambda/b).
\end{eqnarray}
The vertex correction ${\bm\Upsilon}(\Lambda)$ is obtained by integrating modes with $0<|{\bm k}|<\Lambda$, and it can be parameterized as ${\bm\Upsilon}(\Lambda)={\bm v}{\hat\Upsilon}_\Lambda$, with the dimensionless part obeying
\begin{equation}
\frac{d\ln {\hat\Upsilon}_\Lambda}{d\ln \Lambda}= C_v.
\end{equation}
The renormalized fermion-electromagnetic field coupling then becomes
\begin{equation}
S_{A{\Psi}}=(1+C_v l){\bm A}\cdot \int_0^{\Lambda/b} \frac{d^2{\bm k}}{(2\pi)^2} \frac{d\omega}{2\pi}{\bm v}_{\bm k} \Psi^\dagger_< \Psi_<.
\end{equation}
After rescaling, using the relation $C_v=-C_1$, which we prove in the following subsection, one obtains
\begin{equation}
S_{A{\Psi}}={\bm A}\cdot \int_0^{\Lambda} \frac{d^2{\bm k}'}{(2\pi)^2} \frac{d\omega'}{2\pi}{\bm v}'_{\bm k} \left(\Psi'\right)^\dagger \Psi',
\end{equation}
which is of the same form as the unrenormalized coupling term, in accordance with Ward identity. 

Consider finally the superfluid stiffness term generated from integrating out fast modes. We use the standard $1/N$ expansion. Each fermion loop contributes a factor $N$, and each vertex contributes $N^{-1/2}$. The leading order diagrams are the ring diagrams in the RPA approximation, which scale as ${\cal O}(N)$. The next order diagrams are the fermion self-energy and vertex corrections, which scale as ${\cal O}(1)$. However since the fermion-electromagnetic field coupling is proportional to the identity matrix in Nambu space, and the fermion-critical mode coupling is proportional to $\tau_x$ or $\tau_z$, the two fermion loops at the two ends of the ring diagram, which involve both couplings, vanish upon tracing over the Pauli matrices. The leading order contributions from the critical modes are thus the fermion self-energy and vertex corrections. A resummation of the corresponding Feynman diagrams gives
\begin{eqnarray}
\delta\kappa=\int_{\Lambda/b}^\Lambda \frac{d^2{\bm k}}{(2\pi)^2} \frac{d\omega}{2\pi} {\rm Tr}\frac{1}{\left[G_{0}^{-1}({\bm k}, \omega)-\delta\Sigma({\bm k}, \omega)\right]^2}  \left( v_x+\delta\Upsilon_x\right)^2,
\end{eqnarray} 
with the renormalized Green's function and vertex 
\begin{eqnarray}
\frac{1}{G_0^{-1}({\bm k}, \omega)-\delta\Sigma({\bm k}, \omega)}
&=&\frac{1}{1-C_1l} \frac{1}{-i\omega+v'_F k_x\tau_z+v'_\Delta k_y\tau_x},\\
v_x+\delta\Upsilon_x&=&v_x(1+C_vl).
\end{eqnarray} 
Using the relation $C_v=-C_1$, one obtains
\begin{eqnarray}
\delta\kappa={\cal A}'\frac{v'_F}{v'_\Delta}\left( \Lambda-\frac{\Lambda}{b} \right).
\end{eqnarray} 
The partition function then obeys
\begin{eqnarray}
Z(\Lambda, v_F, v_\Delta)=Z(\Lambda/b, v'_F, v'_\Delta)e^{\frac{1}{2}\delta\kappa A^2},
\end{eqnarray} 
and consequently the kernel
\begin{eqnarray}
K(\Lambda, v_F, v_\Delta)=K(\Lambda/b, v'_F, v'_\Delta)+\delta\kappa.
\end{eqnarray} 
Expanding the R.H.S. of the above equation for infinitesimal $l=\ln b$ yields the Callan-Symanzik equation 
\begin{equation}
\left[ \Lambda\frac{\partial}{\partial\Lambda}+\beta_F v_F\frac{\partial}{\partial v_F}+\beta_\Delta v_\Delta\frac{\partial}{\partial v_\Delta}-1\right]K(\Lambda, v_F, v_\Delta)=0,
\end{equation} 
where
\begin{eqnarray}
\beta_F &=&\frac{d \ln v_F}{d\ln\Lambda},\\
\beta_\Delta &=&\frac{d \ln v_\Delta}{d\ln\Lambda}.
\end{eqnarray} 
The solution for finite $l$ is of the form
\begin{equation}
K(\Lambda_0, v_F^0, v_\Delta^0)=e^l K(\Lambda_0 e^{-l}, v_F(l), v_\Delta(l)).
\end{equation}
But by dimensional analysis,
\begin{equation}
K(\Lambda, v_F, v_\Delta)=\Lambda f(v_F/v_\Delta).
\end{equation}
Restoring the band index, we have
\begin{equation}
\frac{d\ln K}{dl}=-1+\frac{d\ln R_v}{dl},
\end{equation}
where $R_v=\sum_a(v_{F a}/v_{\Delta a})$. This equation, together with the RG equations for
the velocity ratios and the coupling strengths 
\begin{eqnarray}
\frac{d\ln(v_{Fa}/v_{\Delta a})}{dl}&=&C_{3a}-C_{2a},\\
\frac{d\ln g_a}{dl}&=&C_{4a},
\end{eqnarray}
fully determine the RG flow of the electromagnetic kernel, and hence the temperature dependence of the penetration depth at low temperatures.

\subsection*{RG coefficients}
The RG equations for the multiband models obviously involve more variables than that of the single band model and hence are more complicated. However we note that near the fixed point, the behavior of the system simplifies significantly due to the enlarged symmetry at the fixed point. Let us consider for example $\tau_z$-coupling with $M=3$ bands, which is relevant for CeCoIn$_5$. By solving the RG equations numerically, one can show that the fixed point is at $v_{Fa}^*/v_{\Delta a}^*=0$, and $g_1^*=g_2^*=g_3^*$. The fixed point has an enlarged $O(3N)$ symmetry. The asymptotic behavior of such a multiband system near the fixed point is thus the same as that of the single band system, with the number of fermion flavors changed from $N$ to $3N$. Hence in real calculations, it is sufficient to retain only the RG coefficients in the single band model, which are included below.

For $\tau_x$-coupling, the RG coefficients $C_{1, 2, 3}$ have been calculated explicitly in  [\onlinecite{Huh08}], 
\begin{eqnarray}
C_1=\frac{2(v_\Delta/v_F)}{\pi^3N}\int_{-\infty}^\infty dx\int_0^{2\pi}d\theta\frac{x^2-\cos^2\theta-(v_\Delta/v_F)^2\sin^2\theta}{[x^2+\cos^2\theta+(v_\Delta/v_F)^2\sin^2\theta]^2}Y_1(x,\theta),\label{EqC1}\\
C_2=\frac{2(v_\Delta/v_F)}{\pi^3N}\int_{-\infty}^\infty dx\int_0^{2\pi}d\theta\frac{-x^2+\cos^2\theta-(v_\Delta/v_F)^2\sin^2\theta}{[x^2+\cos^2\theta+(v_\Delta/v_F)^2\sin^2\theta]^2}Y_1(x,\theta),\label{EqC2}\\
C_3=\frac{2(v_\Delta/v_F)}{\pi^3N}\int_{-\infty}^\infty dx\int_0^{2\pi}d\theta\frac{x^2+\cos^2\theta-(v_\Delta/v_F)^2\sin^2\theta}{[x^2+\cos^2\theta+(v_\Delta/v_F)^2\sin^2\theta]^2}Y_1(x,\theta)\label{EqC3},
\end{eqnarray}
with 
\begin{equation}
Y_1^{-1}(x,\theta)\equiv\frac{x^2+\cos^2\theta}{\sqrt{x^2+\cos^2\theta+(v_\Delta/v_F)^2\sin^2\theta}}+\frac{x^2+\sin^2\theta}{\sqrt{x^2+\sin^2\theta+(v_\Delta/v_F)^2\cos^2\theta}}.
\end{equation}
The results for $\tau_z$-coupling can be obtained by interchanging the two velocities, and $C_2$ with $C_3$ in the above equations, i.e.
\begin{eqnarray}
C_1=f_1(v_\Delta, v_F)\to C_1=f_1(v_F, v_\Delta),\\
C_2=f_2(v_\Delta, v_F)\to C_2=f_3(v_F, v_\Delta),\\
C_3=f_3(v_\Delta, v_F)\to C_3=f_2(v_F, v_\Delta).
\end{eqnarray}
We note from the above expressions the crucial difference between the two cases, namely $C_2-C_3<0$ for $\tau_x$-coupling, and $C_2-C_3>0$ for $\tau_z$-coupling, which gives rise to totally different behavior of the RG flow. For $\tau_x$-coupling, $v_\Delta/v_F$ decreases under the RG flow, and the RG fixed point is at $v_\Delta/v_F=0$ [\onlinecite{Kim08, Huh08}]. For $\tau_z$-coupling, $v_\Delta/v_F$ increases under the RG flow, and the RG fixed point is at $v_F/v_\Delta=0$.

The RG coefficient $C_v$ for the current vertex can be obtained similarly. The one-loop correction to the current vertex is 
\begin{equation}
{\bm \Upsilon}_\Lambda\simeq {\bm v}+\int^{\Lambda}d^2{\bm p}d\Omega {\bm v} \Gamma_1 G_A^{(0)}({\bm p}, \Omega)\tau_0 G_A^{(0)}({\bm p}, \Omega)\Gamma_1 D({\bm p}, \Omega),
\end{equation}
with $D$ the Green's function of the order parameter. Cutoff dependence can be introduced by multiplying scale dependent factors to Green's functions, and the result reads
\begin{equation}
\frac{d{\hat \Upsilon}_\Lambda}{d\ln\Lambda}=\frac{v_F}{8\pi^3}\int_{-\infty}^\infty dx \int_0^{2\pi}d\theta D({\hat{\bm p}}, {\hat\Omega}) \left[\Gamma_1G_A^{(0)}({\hat{\bm p}}, {\hat\Omega})G_A^{(0)}({\hat{\bm p}}, {\hat\Omega})\Gamma_1\right],
\end{equation}
with ${\hat{\bm p}}=(\cos\theta, \sin\theta)$ and ${\hat\Omega}=v_Fx$. Comparing this expression with the corresponding expression for self energy renormalization as obtained in [\onlinecite{Huh08}], one can see that 
\begin{equation}
\frac{d\ln {\hat\Upsilon}_\Lambda}{d\ln \Lambda}\equiv C_v=-C_1=\eta_f.
\label{Ward}
\end{equation}

\twocolumngrid

\bibliographystyle{apsrev4-1}
\bibliography{strings,refs}

\end{document}